 \documentclass[final,3p,times,twocolumn,authoryear]{elsarticle}

\usepackage{amssymb}
\usepackage{changepage}
\usepackage{float}
\usepackage{dsfont}
\usepackage{natbib}
\usepackage{graphicx}
\usepackage{amsmath}
\usepackage{graphics}
\usepackage{multirow}
\usepackage{rotating}
\usepackage{lscape}

\usepackage {comment}
\usepackage [normalem] { ulem }
\usepackage {soul}
\usepackage{xcolor}

\journal{Atmospheric Pollution Research}

\begin{document}

\begin{frontmatter}



\title{Mapping urban air quality using mobile and fixed low cost sensors: a model comparison}

\author[label1,label2]{IDIR Yacine Mohamed}
 \affiliation[label1]{organization={Gustave Eiffel University, PCS-L},
             addressline={25 allée des Marronniers},
             city={Versailles},
             postcode={78000},
             country={France}}

\affiliation[label2]{organization={ESTACA},
            addressline={12 Avenue Paul Delouvrier},
            city={Montigny-le-Bretonneux},
            postcode={78180},
            country={France}}

\author[label3]{ORFILA Olivier}
 \affiliation[label3]{organization={VEDECOM},
             addressline={23 bis allée des Marronniers},
             city={Versailles},
             postcode={78000},
             country={France}}

\author[label4]{CHATELLIER Patrice}
 \affiliation[label4]{organization={Gustave Eiffel University, IMSE},
             addressline={16 Bd Newton},
             city={Champs-sur-Marne},
             postcode={77420},
             country={France}}

\author[label2]{Judalet Vincent}

\author[label5]{GAUFFRE Valentin}
 \affiliation[label5]{organization={AtmoTrack},
             addressline={1 Rue Julien Videment},
             city={Nantes},
             postcode={44200},
             country={France}}

\begin{abstract}

This study addresses the critical challenge of modeling and mapping urban air quality to ascertain pollutant concentrations in unmonitored locations. The advent of low-cost sensors, particularly those deployed in vehicular networks, presents novel datasets that hold the potential to enhance air quality modeling. This research conducts a comprehensive review of ten statistical models drawn from existing literature, using both fixed and mobile low-cost sensor data, alongside ancillary variables, within the urban confines of Nantes, France.

Employing a methodology that includes cross-validation of data from low-cost sensors and validation on fixed air quality monitoring stations, this paper evaluates the models' performance in scenarios of temporal interpolation and prediction. Our findings reveal a pronounced bias in the model outputs when reliant on low-cost sensor data compared to the verification data obtained from fixed stations. Furthermore, machine learning models demonstrated superior performance in predictive scenarios, suggesting their enhanced suitability for forecasting tasks.

The study conclusively indicates that reliance solely on data from low-cost mobile sensors compromises the reliability of air quality models, due to significant accuracy deficiencies. Consequently, we advocate for a directed focus towards the integration and calibration of low-cost sensor data with information from fixed monitoring stations. This approach, rather than an exclusive emphasis on the complexity of statistical modeling techniques, is pivotal for achieving the precision required for effective air quality management and policy-making.\textbf{}

\end{abstract}





\end{frontmatter}


\newpage
\newpage

\section{Introduction}
The harmful effects of air pollution are well known and the World Health Organisation (WHO) is constantly warning of its harmful effects on human health.
According to the WHO database released in April 2022, 99\% of the world's population lives in places where air pollution levels exceed WHO recommended limits \footnote{Source: https://www.who.int/news/item/04-04-2022-billions-of-people-still-breathe-unhealthy-air-new-who-data }. 

 The organization has also declared that air pollution is the leading environmental cause of death, responsible for 7 million premature deaths worldwide in 2018, with 4 million due to outdoor air pollution.\newline

According to \cite{raaschou2013air}, in addition to the adverse effects on vegetation, buildings and visibility, there is obviously a link between particulate matter and the increase in various forms of heart disease and respiratory disorders.

More precisely, there is evidence that PM10 (particulate matter with a diameter of less than 10 $\mu m$) has a short-term exposure effect on respiratory health, and PM2.5 (particulate matter with a diameter of less than 2.5 $\mu m$) are positively associated with increases in mortality due to non-accidental cause, cardiovascular disease and respiratory disease (\cite{lu2015systematic}).\newline

Mitigating air pollution in urban centers and its adverse effects on public health and the environment necessitates a comprehensive, multistage process involving collaboration between scientists and policymakers. Implementing these procedures requires the establishment of obust air quality monitoring systems.\\

The vast majority of health studies \cite{fuller2017air} to date have focused on the relationship between human health effects and longer-term exposures, such as aggregate effects of 1 hour or daily average. These studies investigates the association between pollutant concentrations and health outcomes in locations where direct observations are available, predominantly relying on data from fixed monitoring stations. This approach, however, confines the scope of health impact assessments to areas within the vicinity of these stations, leaving significant urban expanses without precise air quality data. Consequently, the development of accurate pollution maps becomes imperative to bridge this gap, enabling the extension of health studies to encompass broader urban areas, including regions lacking direct observations. Such enhanced spatial coverage is crucial for a comprehensive understanding of the health effects of air pollution across entire cities, thus informing more effective public health policies and interventions.\\

Transportation in urban areas, especially road traffic, is one of the main sources of PM emissions. Cars, buses, and trucks emit PM as well as other pollutants, including nitrogen oxides (NOx) and Volatile Organic Compounds (VOCs) from their tailpipe and non-tailpipe sources, such as tire and brake wear. According to a meta-analysis \cite{heydari2020estimating}, road traffic is responsible on avreage for 27\% of the PM2.5 concentration in urban areas, depending on the city.

Fine particles from road traffic and other sources are highly heterogeneously distributed in the atmosphere and can be found in different concentrations depending on the location [\cite{jeong2019temporal}]. Air pollution levels are also affected by weather conditions, such as wind speed and direction, which can lead to higher or lower concentrations of pollutants in different parts of a city.
Variations in weather conditions lead to a completely heterogeneous distribution of pollutants that can vary from one neighborhood to another
[\cite{berkowicz1996using,scaperdas1999assessing}].\newline

This makes the creation of pollution maps with high spatial accuracy very challenging to produce using only a small number of fixed stations.
However, scientists used different methods to address this problem :
\cite{meng2015land} who successfully modeled NO2 data in Shanghai using a
Land Use Regression (LUR) model, with a R-Squared exceeding 70 \%. \cite{kim2014ordinary} assessed the spatial prediction ability of Oridnary kriging in seven Korean cities for PM10. 
\cite{chen2010land} used a LUR model to predict the spatial concentration distribution of NO2 and PM10 in the region of Tianjin, China.
The resulting maps from these studies are not very precise and reliable at the scale of a city: the number of fixed stations is limited due to various constraints, and the spatio-temporal representation of atmospheric pollution at a fine scale becomes unfeasible, regardless of the model used.


The idea of using new Low Cost Sensors (LCS) developed from recent technological advances, smaller, integrating Global Positioning System (GPS) technology has quickly emerged. This offers scientists additional tools to refine spatial-temporal maps of air pollution, and create new data sets that provide air quality information that was not previously available from the limited number of stationary monitors in the city \cite{schneider2017mapping}.

Numerous instruments from different companies have entered the market \cite{borghi2017miniaturized}. The performance of these sensors is their biggest drawback. Studies have been done to compare its accuracy by field and laboratory evaluations [\cite{feinberg2018long}]. They showed differences in the measurements even between units of the same model.

Several projects have been launched in various cities around the world in an attempt to refine the characterization by overcoming the challenges associated with the use of these sensors [\cite{hasenfratz2015deriving,english2017imperial}].

Due to their small volume, as well as their low energy consumption, low-cost sensors are suitable for deployment in mobile sensor fleets [\cite{zhang2020real}], provided that the problems of sensor communication and sensor network deployment are overcome [\cite{marques2020internet}].

Mobile low-cost sensors have been found to be a useful tool for predicting air quality due to their ability to be easily installed in different locations and to provide real-time data \cite{re2014urban}. This allows a more comprehensive and dynamic understanding of air quality in a given area, as stationary air quality monitoring stations are limited to a single location and may not be able to capture spatial and temporal variations in air pollution \cite{wallace2009mobile}. However, mobile sensors may have their own limitations, such as potential measurement errors \cite{feinberg2018long} and a lack of long-term data which should be taken into account when relying on them to predict air quality.

In this study, we investigate the potential of employing low-cost sensors, encompassing both mobile and stationary configurations, for the modeling of urban air quality. Our objective is to facilitate the generation of detailed urban air pollution maps. Specifically, we address the following research questions:\newline

- What are the most effective statistical models for predicting air quality in urban environments utilizing a network of low-cost sensors ? And which of these models are best suited for forecasting or reconstructing air quality scenarios\newline

- Are the predictions of these models using LCS data in accordance with the measurements of the fixed air quality stations considered as reference ?\newline

To answer those research questions, we need to look at the models already used to map pollution using traditional fixed monitoring stations.
Several studies have been carried out in this field, applying a wide range of models to evaluate the concentration of polutant at unsampled locations. These studies can be classified into four families :

\begin{itemize}

\item[~~~~~~~~] \textbf{Deterministic interpolation methods} Inverse Distance Weighting (IDW) is one of the most applied deterministic interpolation techniques. The weighted average of the data collected from the deployed sensors is used to determine the value at the unsampled location. The closest places are given more weight in this strategy based on the assumption that nearby measurements are more similar in characteristics and that air pollution exhibits localized behavior.
Given its ease of use, this approach is frequently applied as a reference. It was used by \cite{marshall2008within} in Vancouver, Canada, to compare the urban variability of NO and NO2 concentration with a LUR model and an Eulerian grid model.
\cite{wong2004comparison} compared different interpolation methods, including IDW to estimate ozone concentration and PM concentrations.

The limitation of the deterministic interpolation techniques lies in their poor extrapolation accuracy. These techniques are not regarded as models since they don't describe the data and don't indicate the degree of uncertainty in the prediction.

\item[~~~~~~~~] \textbf{Land-Use models}
These models operate under the assumption that local environmental factors, such as land use, weather-related factors, building density and traffic density, are the only factors that affect air quality in a specific location. These models relate the selected environmental predictor factors to the available air quality measurements. Land-use models have two subfamilies, regression models and machine learning.\newline

 1-\textbf{Regression models:}
Regression models are statistical models that take land use as a covariate and assume that observations are independent. They are characterized by their good descriptive power. The main regression models are multiple regression, generalized linear models, and generalized aditive models, although generalized aditive models may be labeled as machine learning in some aspects.

A LUR model developed by \cite{kerckhoffs2015national}, which includes small-scale traffic, large-scale address density, and urban green, explained 71\% the spatial variation for ozone concentrations.
\cite{meng2015land} and  \cite{chen2010land} successfully developed a LUR model for NO2 concentrations in China.

Despite their relatively low complexity, models based on land use regressions can still yield satisfactory results. These models are also capable of providing insights into the ways in which environmental factors impact the concentration of pollutants. However, LUR models are constrained by the amount of information required from other variables or costly to acquire.\newline

2-\textbf{Machine learning:}
When new insights must be gained from vast, varied, and dynamic data sets, machine learning is particularly useful.
A machine learning algorithm analyses the training data and produces an inferred function that can be used for prediction in unsampled locations. Typically, this function is quite complex and cannot be directly interpretable by the user.
Numerous studies applied this method to predict air pollution levels: \cite{singh2013identifying} identified pollution sources and predicted urban air quality using ensemble learning methods.
\cite{cabaneros2019review} provided a review of Artificial Neural Network (ANN) models for the prediction of air pollution. Some machine learning algorithms were combined with fuzzy models to predict levels of air pollution \cite{cacciola2013aspects}.
Machine learning algorithms are viewed as "black boxes" with low descriptive capabilities that struggle to outperform other models with limited data.

\item[~~~~~~~~] \textbf{Geostatistics} Geostatistics predicts values at unsampled locations using a weighted linear (implaying a gaussian process) combination of observations. Weights are determined by the covariance structure of the data inferred from the actual data structure.

 \cite{kim2014ordinary} developed an Ordinary Kriging (OK) prediction model to predict long-term PM concentrations in seven major Korean cities.
\cite{whitworth2011kriged} modeled ambient air levels of benzene in an urban environment.
\cite{schneider2017mapping} modeled nitrogen dioxide concentration in Oslo, using the dispersion model output as a covariable in a geostatistic model. \cite{idir2021mapping} compared several geostatistical models using LCS data.
More sophisticated than IDW and regression modeling, geostatistics also provides the uncertainty associated with the prediction. However, these techniques suffer from a relatively high computational cost.

\item[~~~~~~~~] \textbf{Dispersion models} Dispersion models reproduce how physical and chemical processes lead to the production of atmospheric pollutants. They involve environmental factors, such as those used in LUR models, and have been frequently applied in traffic-related pollution prediction.

\cite{hamer2019urban} described the Eulerian urban dispersion model EPISODE and its application to the modeling of the pollution concentration $NO_2$.
\cite{fallah2017integrating} improved the characterization of near-road air pollution using a regional Gaussian dispersion model.
\cite{gibson2013dispersion} used the AERMOD Gaussian plume air dispersion model to evaluate PM, NOx and $SO_2$.
These techniques have a number of drawbacks, including high processing costs connected to the difficult issue of modeling tiny-scale random fluctuations.

\end{itemize}

\begin{figure*}[h!]
\includegraphics[width=\textwidth,height=10cm]{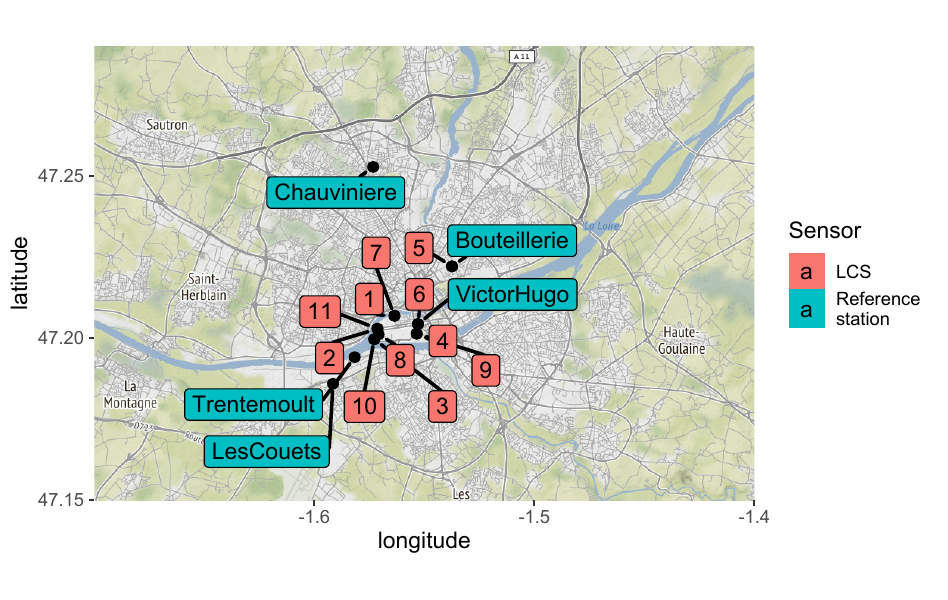}
\caption{Location of fixed LCS and monitoring stations in Nantes}
\label{figure1}
\end{figure*}

Table \ref{table-bib} lists some published studies using low-cost mobile sensors for air quality mapping. As shown in this table, different models from different families were used to tackle the challenge of mapping urban air quality using low cost mobile sensors.


Beyond the creation of maps with a minimum prediction error, the models must be able to obtain information on the pollutant dispersion process. Understanding this process, its origin, and the causes of pollution peaks is essential. This is why models with a good descriptive power will be preferred and will allow the concerned organizations to act as closely as possible and implement efficient measures to fight against air pollution.

Few studies have reviewed this emerging topic. \cite{xie2017review} reviewed  monitoring and exposure assessment methods for air pollution, while \cite{bertero2020urban} compared three different models (LUR, kriging, and neural network) on a synthetic data set. \\

This study extends this work by integrating various models, applying them to a real dataset derived from both fixed and mobile sensor. Through a rigorous comparison across various scenarios, this study employs cross-validation techniques and further validation against data from fixed monitoring stations to assess the reliability and accuracy of low-cost sensors data.
The primary objective of this paper is to improve understanding of the subject and assist users in their pollution mapping efforts based on their equipment and goals. To the best of our knowledge, no prior research has compared all the model families employed in this study.



\begin{table*}[t]
\begin{center}
\resizebox{\textwidth}{!}{%
\begin{tabular}{ |c|c|c|c|c|c| } 
\hline
Article & Method & Region & Pollutants & Carrier \\
\hline
\hline
 \textbf{\cite{marjovi2017extending}} & Regressions & Lausane, Switzerland & UFP & Bus  \\  
\textbf{\cite{hart2020monitoring}} &Regressions & Texas, USA & PM2.5 & Bikes \\  
\textbf{\cite{ apte2017high}} &  Reduction algorithm & Oakland, USA &   NO, NO2,BC & Cars  \\  
\textbf{\cite{ hasenfratz2014pushing}} & Regressions & Zurich, Switzerland & UFP & Trams \\  
\textbf{\cite{hasenfratz2015deriving}} & Regressions & Zurich, Switzerland & UFP & Trams \\  
\textbf{\cite{ marjovi2015high}}  &  Network regression &Lausane, Switzerland   & UFP & Bus \\ 
\textbf{\cite{li2014estimating}} & Geostatistics & Zurich, Switzerland& UFP& Trams\\

 \textbf{\cite{lim2019mapping}} & Machine learning & Seoul, South corea &PM2.5  & Pedestrians \\  
\textbf{\cite{adams2016mapping}} & Neural network & Hamilton, Canada & NO2 & Vans  \\  
\textbf{\cite{hankey2015land}} & Regressions & Minneapolis, USA &BC PM2.5  & Bikes  \\ 
\textbf{\cite{ gressent2020data}}& Geostatistics & Nantes, France &PM10 & Cars \\
\textbf{\cite{do2020graph}}&  Neural network & Anvers, Belgium & Different pollutants & Bikes\\
\textbf{\cite{zhang2020real}} & Machine learning & Songdo, South corea &  CO2,PM2.5,PM10 & Cars\\
\textbf{\cite{song2020deep}} &Machine learning &  Beijing, China &  PM2.5 & Cars \\
\textbf{\cite{van2020development}} &Regressions &  Gand, Belgium &  BC & Bikes \\
\textbf{\cite{guan2020fine}} &Geostatistics &  Oakland, USA &  NO2 & Cars \\
\textbf{\cite{mariano2020pollution}} & Machine learning & Zurich, Switzerland  & UFP  &Trams\\
\textbf{\cite{ma2020fine}} & Machine learning & China  & PM2.5& Cars \\
\textbf{\cite{idir2021mapping}} & Geostatistics & Zurich, Switzerland  & O3& Tram \\

\textbf{\cite{zhao2021urban}} & Machine learning & China  & PM2.5& Cars \\
\textbf{\cite{qin2021street}} & Random forest with coordinates & China  & PM2.5& Cars \\
\textbf{\cite{qin2022fine}} &Deep forest & Antwerp ,Belgium  & NO2& Vans \\

\hline
\end{tabular}}
\end{center}
\label{table-bib}
\caption{Maping Air quality studies using mobile sensors.}
\end{table*}

\section{Data methodology and models}
This section begins with a description of the dataset used, followed by the methodology for model evaluation. It concludes with an overview of the models applied to the dataset.
\subsection{Data}
This study analyzed a diverse range of data, including pollution data obtained from low-cost sensors and stationary air quality monitoring stations. Additionally, supplementary data about the surrounding environment of pollution observations was also considered.
The objective of this subsection is to clarify these variables, their origins, and includes examples for enhanced comprehension.

\subsubsection{Pollution data}
The pollution data used in this article were collected from three distinct sources:\newline
1- The primary air quality data were obtained from low-cost sensors managed by Atmotrack.\newline
2- Air quality data from fixed air quality measurement stations located in Nantes, France were utilized as reference. \newline
3- Air quality data from the CHIMERE chemistry-transport model was used as a covariate.\newline

Atmotrack ( atmotrack.fr) is a company specialized in air quality measurement based in Nantes, France.
It has a fleet of low-cost mobile sensors placed on utility vehicles such as those used by the post office, as well as fixed low-cost sensors placed at strategic locations in the city of Nantes, notably near fixed air quality stations .

The dataset provided by Atmotrack is diverse and contains various types of measurements of PM2.5 and PM10. These measurements were obtained using low-cost, mobile, and fixed sensors and collected throughout the month of November 2018. It is worth noting that PM10 observations are more frequent at fixed air quality measurement stations. Therefore,PM10 was selected as the initial pollutant to investigate in this study. 

The initial 20 days of the data set primarily comprise data obtained from low-cost fixed sensors. To ensure homogeneity in our dataset and a sufficient amount of data to train our models, we have limited ourselves to the final 10 days of November 2018, from 20 to 30 November 2018.

The majority of the observations are concentrated in and around the city of Nantes. However, a small portion of the data obtained from the mobile sensors, which were not present in this area, have been excluded from further analysis. Figure \ref{figure1} depicts the study area that has been retained.

Out of the 15 fixed low-cost sensors, three of them were situated in close proximity to the "Bouteillerie" fixed monitoring station, while three others were located near the "Victor Hugo" station. Figure \ref{figure1} provides a visual representation of the location of both the fixed air quality stations and the low-cost fixed sensors.

To eliminate redundancy in the data, only one set of data from each location with multiple sensors was considered.

Since mobile sensors are installed on top of utility cars using opportunistic sampling, the amount of data collected from these sensors tends to decrease during weeknights and cease on Sundays. To maintain consistency between the data collected from both fixed and mobile sensors, our study will only use data collected between 4am and 7pm from Monday to Saturday, inclusive.

The low-cost sensors were precalibrated with the fixed stations by Atmotrack. To maintain the integrity of the data, a minor post-processing procedure was carried out. Specifically, the first five minutes of mobile runs were excluded from the data set, as it was observed that during this period the mobile sensors produced distorted data. Following this, a running median of 15 observations per sensor was applied to smouth out sensor errors.

The dataset comprises a total of 132,891 observations, with 81,461 observations collected from mobile low-cost sensors and 51,430 observations from fixed low-cost sensors. Further information regarding the Atmotrack dataset can be found in \cite{gressent2020data}.\\

In addition to the low-cost data from atmotrack, we have access to the data from the monitoring fixed station managed by Air Pays-de-la-Loire and available at "data.airpl.org". It consists of a series of measurements from five reference stations for the whole month of November at a frequency of one observation per 15 min.
The location of Atmotrack's low-cost fixed sensors and the fixed monitoring station can be found in Figure~\ref{figure1}.

\begin{figure*}[t]
\includegraphics[width=\textwidth]
{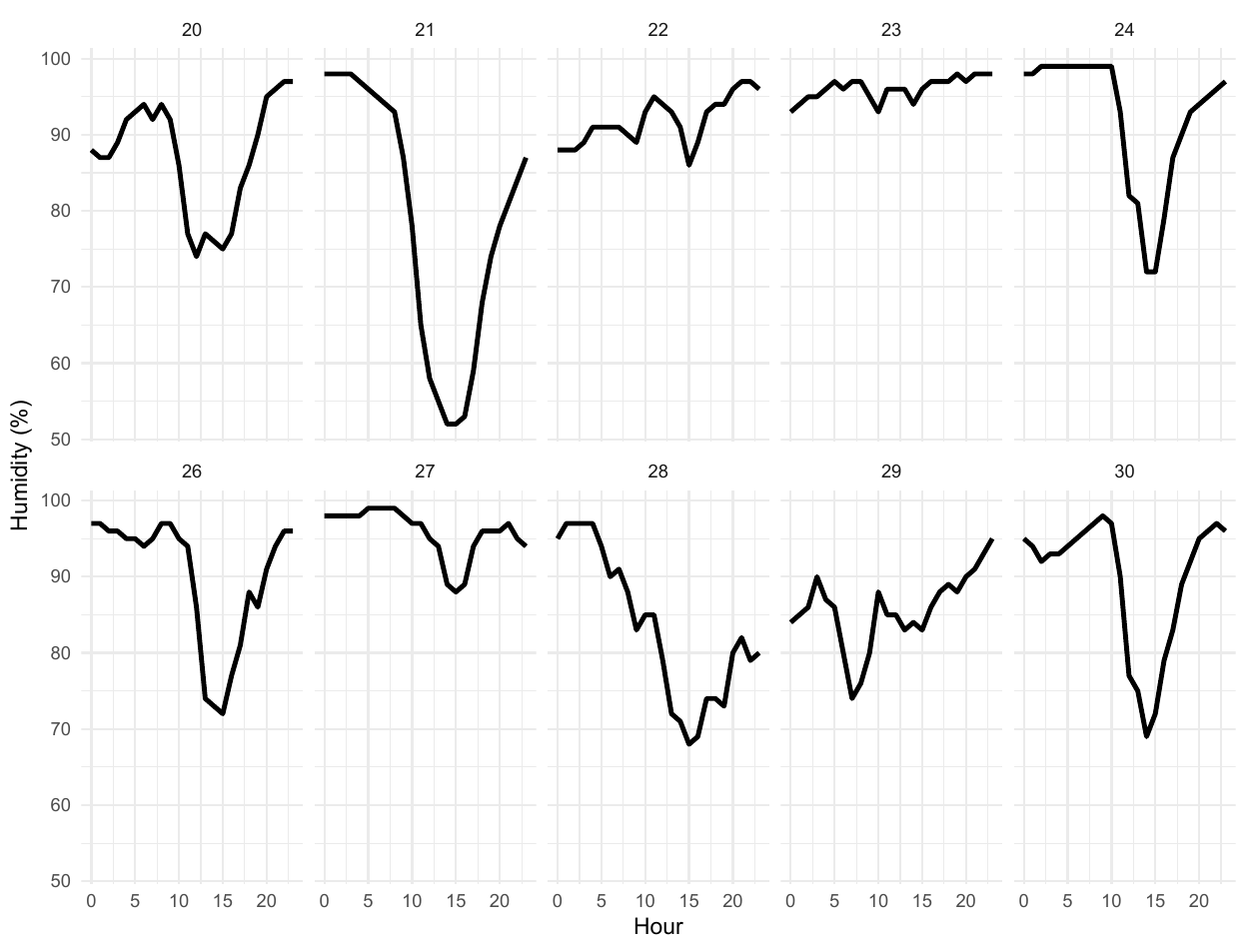}
\caption{Humidity data in Nantes city} 
\label{figure2}
\end{figure*}

\subsubsection{Auxilary data}
Different studies have used a variety of covariate categories, such as data from fixed monitoring stations in \cite{idir2021mapping}, annual average pollution data in \cite{gressent2020data}, or spatial land use covariates in \cite{hasenfratz2015deriving}.

In addition to the pollution data collected by low-cost (mobile and fixed) sensors and fixed air quality stations, additional covariates are required to establish a link between pollution measurement and various environmental characteristics. After investigating this link, these covariates enable the model to predict the concentration of pollutants in areas where no pollution measurement has been taken. As such, unlike pollution data, these data must be accessible throughout the study area, including areas not sampled by pollution sensors.
Several types of covariates have been used in this study, divided into three groups :

\begin{itemize}
\item[~~~~~~~~] \textbf{Temporal covariates}

Temporal covariates are variables that vary only with time and not with space. In other words, if two observations are taken at different locations within the city but at the same time, they would have the same value for these covariates. In this study, meteorological data, specifically temperature and humidity, were used as temporal covariates. Although these variables are not evenly distributed across the entire city, a more precise dataset reflecting this spatial variation is unavailable. Understanding their overall temporal trend can help predict trends in pollutant concentrations. The meteorological data was obtained from the website "public.opendatasoft.com".

Figure \ref{figure2} displays the humidity data for the study period.

\item[~~~~~~~~] \textbf{Spatial covariates}\\
Unlike temporal covariates that vary only in time, spatial covariates vary only in space. This means that if two observations are made at the same location, regardless of when they were collected, they will have the same value for these covariates. These covariates are the same for low-cost fixed sensors and fixed air quality stations. Most of these covariates were obtained from the Open Street Map. Since road traffic is an important factor that affects PM concentration, we first extracted the proximity of the various types of roads present in our study area. We also calculated the number of buildings present in a buffer zone of 500 meters radius. The Nantes digital field model was used to determine the altitude or elevation of each observation. Figure \ref{figure3} illustrates some interesting spatial covariates.

\begin{figure*}[h!]
\includegraphics[width=\textwidth]{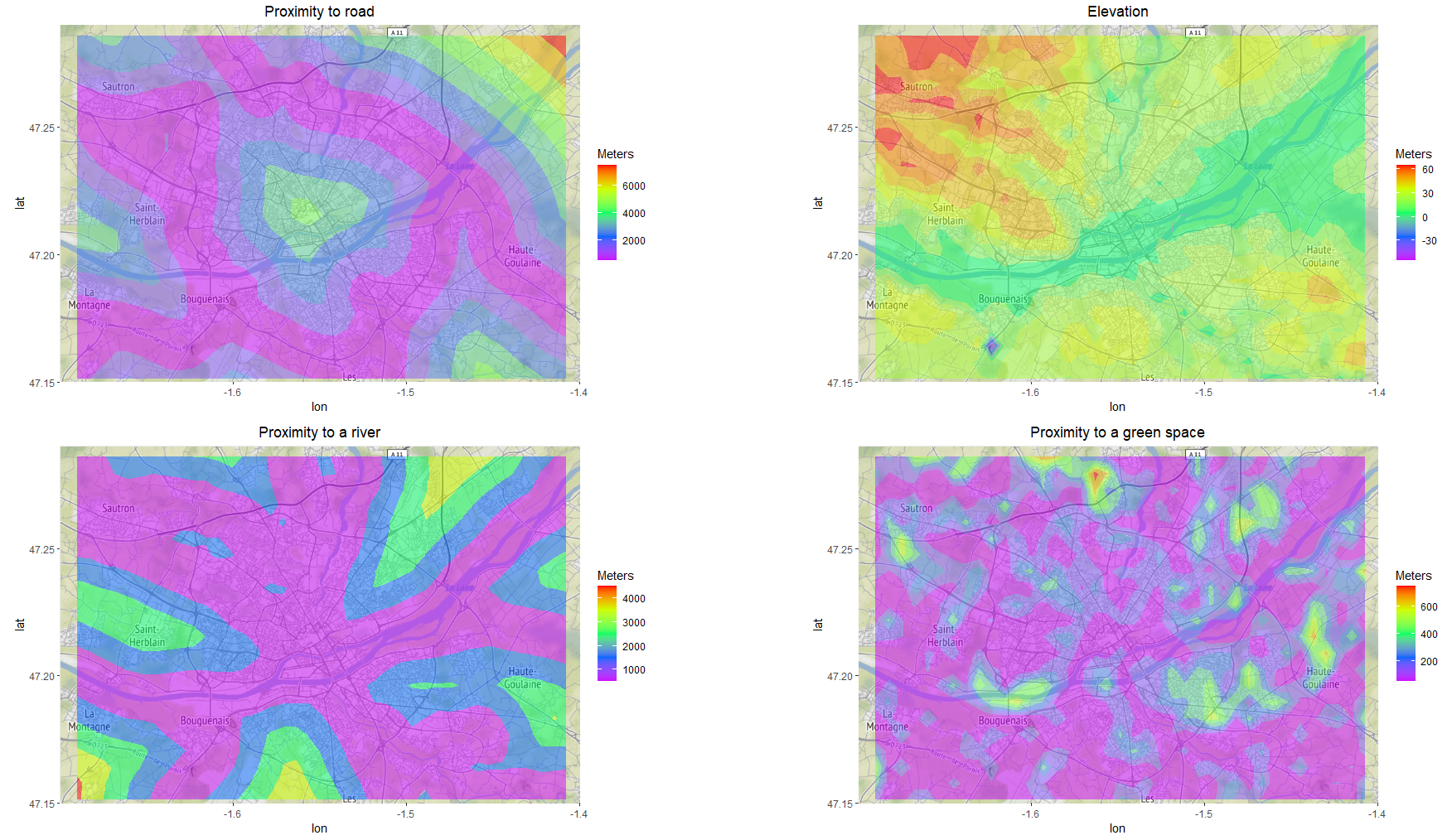}
\caption{Illustration of some spatial covrariates used in this study.}
\label{figure3}
\end{figure*}

\item[~~~~~~~~] \textbf{Spatio-temporal covariates}

CHIMERE is a three-dimensional Chemistry Transport Model (CTM), based on the integration of the mass continuity equation for the concentrations of several chemical species.
This model is used by a large number of air quality monitoring agencies in Europe and is the most used in France.
The regional air quality production of the Copernicus Atmosphere Monitoring Service (CAMS) provides 4-day, daily forecasts of the main atmospheric pollutants concentrations, in the lowest layers of the atmosphere for the European domain.  Their horizontal coverage is 0.1 ° (approximatively 10 to 20 km).
Alongside forecasts, the models perform daily analyses of pollutants near the surface by assimilating, i.e. merging, 1-day old observations with the models. 
Analysis and the 1-day forecast have been selected as relevant covariates. These two spatio-temporal covaraites are available at hourly time steps. 
There is no large spatial variance on the scale of our study area. Figure \ref{CHIMERE} shows the temporal variation of these two covariates.

\begin{figure*}[h!]
\includegraphics[width=\textwidth]{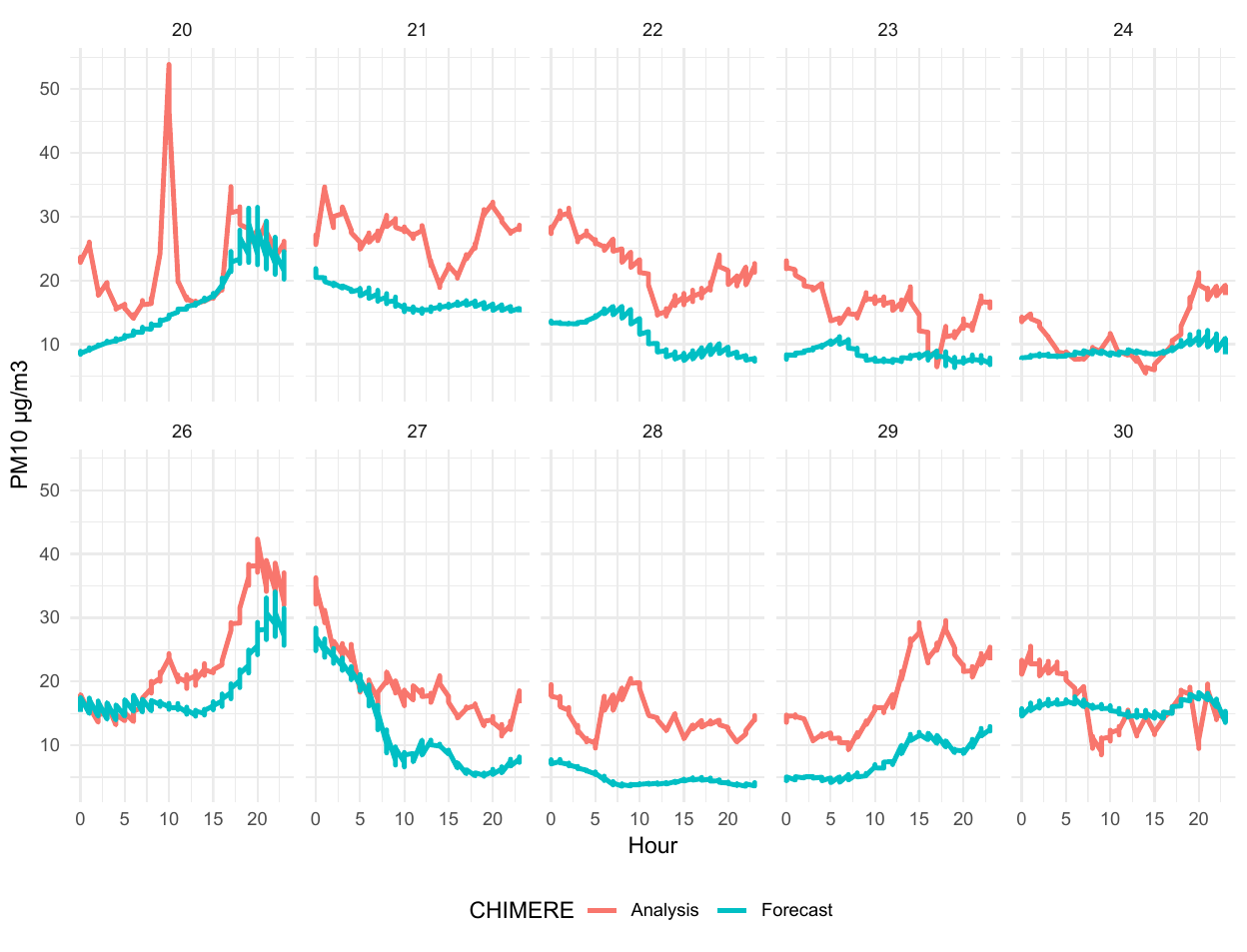}
\caption{Temporal series of CHIMERE forecast and analysis} 
\label{CHIMERE}
\end{figure*}

\end{itemize}

To ensure that each low-cost sensor observation is associated with the corresponding covariates, interpolation is required because some LCS samples are more frequently than 1Hz and some covariates only have data every 15 minutes or an hour. Table \ref{tab:covariate} provides a summary of all the covariates used in this study, including their source and dimension.

\begin{table}[h!]
\resizebox{\columnwidth}{!}{%
\begin{tabular}{ |c|c|c|c|c|c| } 
\hline
Variable & Origin & Dimension\\
\hline
\hline
 Temperature & public.opendatasoft &Temporal \\  
Humidity&public.opendatasoft &Temporal\\  
 Elevation& data.nantesmetropole &Spatial \\  
Proximity to : motorway&Open Street Map&Spatial\\  
Proximity to : trunk& Open Street Map&Spatial\\  
 Proximity to : primary&  Open Street Map &Spatial\\  
 Proximity to : secondary&Open Street Map&Spatial\\  
Proximity to : tertiary&Open Street Map&Spatial \\  
 Proximity to : residential&Open Street Map&Spatial\\  
 Proximité d'un espace vert&Open Street Map&Spatial\\  
Proximity to industrial site&Open Street Map&Spatial\\  
 Nb of buildings in 500m&Open Street Map&Spatial\\  
  Proximity to a river&Open Street Map&Spatial\\  
 CHIMERE analysis& CAMS  &Spatio-temporal\\  
 CHIMERE forecast&CAMS &Spatio-temporal\\

\hline
\end{tabular}}
\caption{Summary of covariates used in this study.}
\label{tab:covariate}
\end{table}

\subsection{Methodology}
This section outlines the methodology applied to address our research questions. The primary research questions are as follows: "What statistical models are more efficient in mapping urban air quality using fixed and mobile LCSs, considering prediction accuracy in terms of temporal interpolation or forecasting? Can LCSs substitute traditional air quality measurement networks?"
To answer those research questions, a set of candidates models has been selected from the literature and compared on the basis of predefined performance indicators estimated from an experimental dataset. The remainder of this section details these methodological steps.

\subsubsection{Models selection}
Models used in this article have been selected from the literature in order to represent a wide range of modeling principles, from statistal models to neural networks. A selection criterion was their proven ability to work with spatio temporal data and especially with mobile sensors.

\subsubsection{Models Validation Process}

To asses the numerical performances of the differents models, we compare them in two different scenarios :
\begin{itemize}
\item Temporal interpolation: The objective is to recreate a picture of pollution during a period in which we have observations. In practice we will use the data of day D to predict the pollution on day D.
\item Forecasting : This involves predicting pollution levels for a period in which no observations are available using previous observations. In practice, we will use the data of day D to predict the pollution on day D+1.
\end{itemize}

In each of the two scenarios, two types of evaluations will be conducted:
\begin{itemize}
\item Cross-validation on low cost sensors: For the first scenario, a leave-one-out cross-validation will be performed on N low-cost sensors. N-1 sensors will be selected to train the models and validated on the remaining sensor, by varying the validation sensor each time. For the second scenario, the N LCS sensors deployed on day D will be used to predict and validate on the N' sensors deployed on day D+1.
\item Validation on fixed air quality stations: Fixed stations are considered as a reference (true signal). Although there is no direct connection between their data and the data from low-cost sensors, the latter measure the same quantities. This evaluation also aims to evaluate the use of low-cost sensors in general, in addition to the models, their calibration, and their potential replacement of the classic fixed station network. In the first scenario, the models will be trained on data from low-cost sensors of day D to predict on the five fixed stations of day D. In the second scenario, the model will be trained on low-cost sensor data of day D to predict the value of the five fixed air quality stations on day D+1.
\end{itemize}

In both scenarios, for both validation cases and for all models examined, four performance indicators are computed to assess the model's ability to accurately predict air quality.

In addition, at the end of the article, a comparison of the maps produced by all models, utilizing the data from the most frequently sampled day, will be presented.

\subsubsection{Performance indicator}
We compared ten models from the literature, using data from LCS and fixed stations. We evaluate the outcome of each procedure using the following four numerical performance indicators.\\

The Root Mean Squared Error (RMSE) is widely accepted as the primary performance indicator for assessing the deviation between the predicted values of a model or estimator and the actual observed values. Minimizing this metric is a common objective when building models.

\begin{equation} \label{RMSE}
RMSE = \sqrt{\frac{\sum_{i=1}^{n}(Z^*_i-Z_i)^2}{n}}
\end{equation}

To monitor the unbiasedness of the estimators, the bias performance indicator was adopted.
\begin{equation} \label{BIAS}
BIAS =  1/n\sum_{i=1}^{n}(Z^*_i-Z_i)
\end{equation}

To deal with low-cost sensors, the correlation performance indicator was chosen. Measuring correlation performance and comparing it to the RMSE is necessary in the case of constant bias.
\begin{equation} \label{COR}
CORR= \frac{\sum_{i=1}^{n}(Z^*_i-\bar{Z^*})(Z_i-\bar{Z})}{\sqrt{  \sum_{i=1}^{n}(Z^*_i\bar{Z^*})^2\sum_{i=1}^{n}(Z_i-\bar{Z})^2      }  }
\end{equation}

The Maximum Absolute Error (MAE) has been selected as a means to identify models that may exhibit significant bias only under certain conditions.

\begin{equation} \label{MAE}
MAE= max(|Z^*-Z|)
\end{equation}

\subsection{Models} 
In this section we will discuss in detail the models used and the parameters applied.
To follow the bibliography and deal with the lognormal distribution of the pollution data, the concentration measurements were log transformed before applaying any model.
In the following $Z_i$ refers to the ith LCS (already log-transformed) observation, embedded in the spatio-temporal location i, $X_i$ to the covariate associated with the observation $i$, $P$ to the dimenstion of $X$ and $n$ to the size of vector $Z$.
All the computations were made in R.
\subsubsection{Inverse Distance Weighting}

Inverse Distance Weighting (IDW) is a type of deterministic method that assigns values to non-sampled places $Z'$ using a linear combination of values from sampled points weighted by the inverse distance.

The general formula for the IDW is given by Equation~(\ref{eq:15}):
\vspace{24pt}
\begin{equation} \label{eq:15}
Z'=\sum_{i=1}^n \lambda_i Z_i
\end{equation}
with:
\begin{equation} \label{eq:16}
\lambda_i=\frac{1/d_i^p}{\sum_{i=1}^{n} {1/d_i^p} } 
\end{equation}

$ d_i $ represents the distance between $ Z' $ and $ Z_i $. Weights decrease as distance increases, especially as the power value $p$ increases. As with the previous methods, points in the neighborhood have a heavier weight and have more influence on the prediction, thus, resulting in a local spatio-temporal interpolation.
In this study, the following definition of a spatio-temporal distance was chosen:
\begin{equation} \label{eq:17}
d_{i}=\sqrt{(x_{i}-x')^2+(y_{i}-y')^2+C\cdot (t_{i}-t')^2}
\end{equation}

The parameter $p$ was fixed at 2, while $C$ was obtained by cross-validation.

\subsubsection{Linear regression}
The most commonly used statistical tool is Linear Regression (LR), which is characterized by its simplicity, low complexity, and high descriptive power. However, compared to the IDW method, it requires a large number of covariates. The linear regression formula (\ref{linear regression}) assumes a linear relationship between the observations and the covariates.

\begin{equation} \label{linear regression}
Z_i=\beta X_i + \epsilon_i
\end{equation}

The Beta's are estimated by ordinary least squares or by maximum likelihood under the assumption that $\epsilon$ (the random part of Z) are independent and follow a normal distribution, so that $Z$ also follows a normal distribution. The term $\beta X$ represents the overall mean (trend) of $Z$.
Linear regression assumes that the concentration of a pollutant only depends on a linear relationship with its covariates and that the observations are independent of each other. However, this assumption is not valid in the case of pollution because dispersion processes are known to exist, violating the independence assumption. However, despite this limitation, a linear regression model can still capture a significant portion of the variance in the data while maintaining a certain level of simplicity.

\subsubsection{Network Regression } 

Following \cite{ marjovi2015high} Network Regression (NR) is very similar to linear regression. It also assumes that the covariates have a linear impact on the concentration of pollutants. In order to alleviate the problem of the independence of the observations between them, another covariate, which represents the average of the concentration of PM of the similar streets, is added to the equation.

\begin{equation} \label{eq:network regression}
Z_i= \beta X_i + \epsilon_i +\beta_{p+1}  \sum_{[i-j]\in E}^{n'} Z_j/n'
\end{equation}

The last term in equation (\ref{eq:network regression}) indicates that we have used the avreage observations of the street segments connected ( in the graph or network $E$) to the observation segment i, as an input to estimate the concentration of polution of the street segment i. All the difficulty lies in determining the correlated streets, that is, constructing the graph $E$.
\cite{marjovi2015high} defined the distance between the correlated streets as the similarity of the covariates \ref{eq:network regression distence}.

\begin{equation} \label{eq:network regression distence}
d(i,j)=  \left\lVert X_i-X_j \right\rVert
\end{equation}

To establish connectivity among street segments, we calculate their pairwise distances and link each segment with its closest neighbor. We continue this process, connecting the nodes based on their minimal covariates distance, until the entire graph $E$ becomes fully connected.

\subsubsection{General Aditive Model }
Used by \cite{hasenfratz2015deriving}, the General Aditive Model (GAM) assumes that observations are independent and that there is a nonlinear relationship between observations and covariates:

\begin{equation} \label{eq:2}
Z_i= S( X_i) + \epsilon_i
\end{equation}

where S denotes a smooth nonlinear function of the covariates Xi. In this paper, S is a smooth regression spline with an upper bound of 9 degrees of freedom.

 General aditive models are at the frontier between regression models and machine learning models.

\subsubsection{Geostatistics : Gaussian process}
 Geostatistical techniques rely on statistical models that use random functions to interpolate and extrapolate data across a spatial or spatio-temporal domain. These models can also incorporate other data. In our case, the random function is a Gaussian process.

The observations $Z$ of the LCS sensors are modeled as a noisy realisation of the true porcces $\mu$ so that 
\begin{equation}\label{gaussian process}
    Z/\mu \sim \mathcal{N}(\mu,\,\tau^{2}I)
\end{equation}
The true process is decomposed into two parts, one deterministic, the other random : 
\begin{equation}\label{decomposition}
 \mu(x,t) =  \beta X + \nu(x,t)
\end{equation}
The model allows the global trend to be captured by the deterministic part using a linear regression, while the remain variability is captured by the small scale process $\nu(x,t)$.

The small-scale process follows a Gaussian process with
mean zero and an isotropic spatio-temporal covariance function $ R(\theta)$.

\begin{equation} \label{modele geostat}
Z \sim \mathcal{N}(\beta X,R(\theta)+\tau^{2}I)
\end{equation}
Two ways of estimating the parameters of the model described by equation~(\ref{modele geostat}) will be compared :

\begin{itemize}
\item Maximum likelihood :
Calculating the likelihood of the geostatistical model is computationally infeasible for large data sets such as the one used in this study.
This paper uses the Vecchia approximation to likelihood (\cite{vecchia1988estimation}; \cite{katzfuss2020vecchia})
resulting in the Gaussian process with maximum likelihood algorithm (GP\textunderscore ML).

\item Spatiotemporal variograms ( GP\textunderscore VG) :
Another way to estimate the parameters is by means of spatiotemporal variograms like those used in \cite{idir2021mapping}.
The parameters of the mean $\beta $ are estimated by ordinary least squares; the rest of the parameters of equation (\ref{modele geostat}) $(\theta, \tau)$ are estimated by weighted least squares on the experimental variogram \cite{pebesma2016spatio}.

\end{itemize}


\subsubsection{Random forest}
The  Random Forest (RF) algorithm combines Ensemble learning methods with the decision tree framework to create multiple randomly drawn decision trees from the data, averaging the results to output a new result that often leads to strong predictions.
It usually performs great on many problems, including features with nonlinear relationships. Disadvantages, however, include the following: there is no interpretability, overfitting may easily occur, and one must choose the number of trees to include in the model.

The core algorithm for building decision trees called ID3 by \cite{quinlan1986induction} employs a greedy top-down search through the space of possible branches without backtracking. The ID3 algorithm can be used to construct a regression decision tree by replacing the information gain with the standard deviation reduction.

The standard deviation reduction is based on the decrease in standard deviation after a data set is split into an attribute. Constructing a decision tree is all about finding the attributes that return the highest standard deviation reduction (i.e., the most homogeneous branches).	

\subsubsection{XGBoost}
At a high level, XGBoost works by sequentially building a model to predict the target variable (PM concentration in our case) based on the input features (covariates). This is done by fitting weak learners (for example, decision trees) to the negative gradient of a loss function. The loss function measures the difference between the predicted value and the true value. In each iteration, the algorithm fits a new weak learner to the residual error of the previous iteration, with the goal of minimizing the overall loss. The final model is a weighted sum of all weak learners, where the weights are determined by the optimization process.

XGBoost is implemented in an open source software library that provides a gradient boosting framework for C++, Java, Python, R, and Julia. It is an implementation of gradient-boosted decision trees designed for speed and performance that is predominant in machine learning competitions.

\subsubsection{Support vector machine}  

\cite{cortes1995support} proposed the Support Vector Machine (SVM) method for classification and regression. The regression used in this method can be considered as a nonparametric technique because it relies on kernel functions.

In this paper, we use the radian kernel and implement linear epsilon-insensitive SVM ($\epsilon$-SVM) regression.
In $\epsilon$-SVM regression, the goal is to find a function $f(X)$ that deviates from $Z$ by a value no greater than $\epsilon$ for each training point $X_i$, and at the same time is as flat as possible:
\begin{equation}\label{SVM equation}
    Z_i=f(X_i)+\epsilon
\end{equation}
    
To find the linear function $f(Z_i)= \beta X_i + b$ and ensure that it is as flat as possible, we formulated a convex optimization problem to minimize the norm value for the parameter vector $\beta$.

Analogously to the “soft margin” loss function, one can introduce slack variables to deal with otherwise infeasible 
constraints of the optimization problem . Hence we arrive at
the formulation stated in \cite{cortes1995support}.


A constant ($C > 0$) determines the trade-off between the flatness of $f$ and the amount up to which deviations greater than $\epsilon$ are tolerated.\\
The next step is to make the SV algorithm non-linear. This is achieved by preprocessing the training patterns $X$ using a map: $X \longrightarrow F$ in some feature space F, as described in \cite{aizerman1964probability} and \cite{nilsson1965learning} and then applying the standard SV regression algorithm.

\subsubsection{Artificial neutral network}

In an artificial neutral network (ANN), a transfer function translates the input signals to output signals. Four types of transfer functions are commonly used: unit step (threshold), sigmoid, piecewise linear, and Gaussian.
A feed-forward network is a nonrecurrent network which contains inputs, outputs, and hidden layers; the signals can only travel in one direction. The input data is passed onto a layer of processing elements, where it performs calculations. Each processing element makes its computation based on a weighted sum of its inputs.


A k-layer neural network is a mathematical function $f$, which is a composition of multivariate functions: $f_1$, $f_2$, \dots, $f_k$, and $g$, defined as:
\begin{equation}\label{ANN1}
 f:\mathbb{R}^n \longrightarrow \mathbb{R}^p
\end{equation}

\begin{equation}\label{ANN2}
f(X)= g\circ f_k \circ ... \circ f_1 (X)  
\end{equation}

\begin{equation}\label{ANN3}
 f_i(X)=  a (w_ix+b_i) 
\end{equation}

$n$ is the dimension of the input $X$ ( 12 in our study).

Each function $f_i$ is a composed function with the functions :
$wX+b$ a linear combination of the input x with its coefficients w, plus a bias b.

$a$ is called an activation function (it is the same for each intermediary function $f_i$)
The activation function can take various forms, in our case we used the sigmoid function.



\section{Results}
This section shows different results from the application of the methodology to the dataset. First, we present the detailed results obtained from some performance indicators for prediction and interpolation scenarios. The validation will be done with LCS and fixed monitoring stations.
The next section presents a table summarizing all the results obtained for a set of performance indicators, scenarios, validation methods, and models.
Finally, we will examine and compare the PM concentration maps produced by the 10 different models for the most sampled day.

\subsection{Performence results}
\subsubsection{Cross-validation on LCS data}

\begin{figure}[H]

\includegraphics[width=\columnwidth]{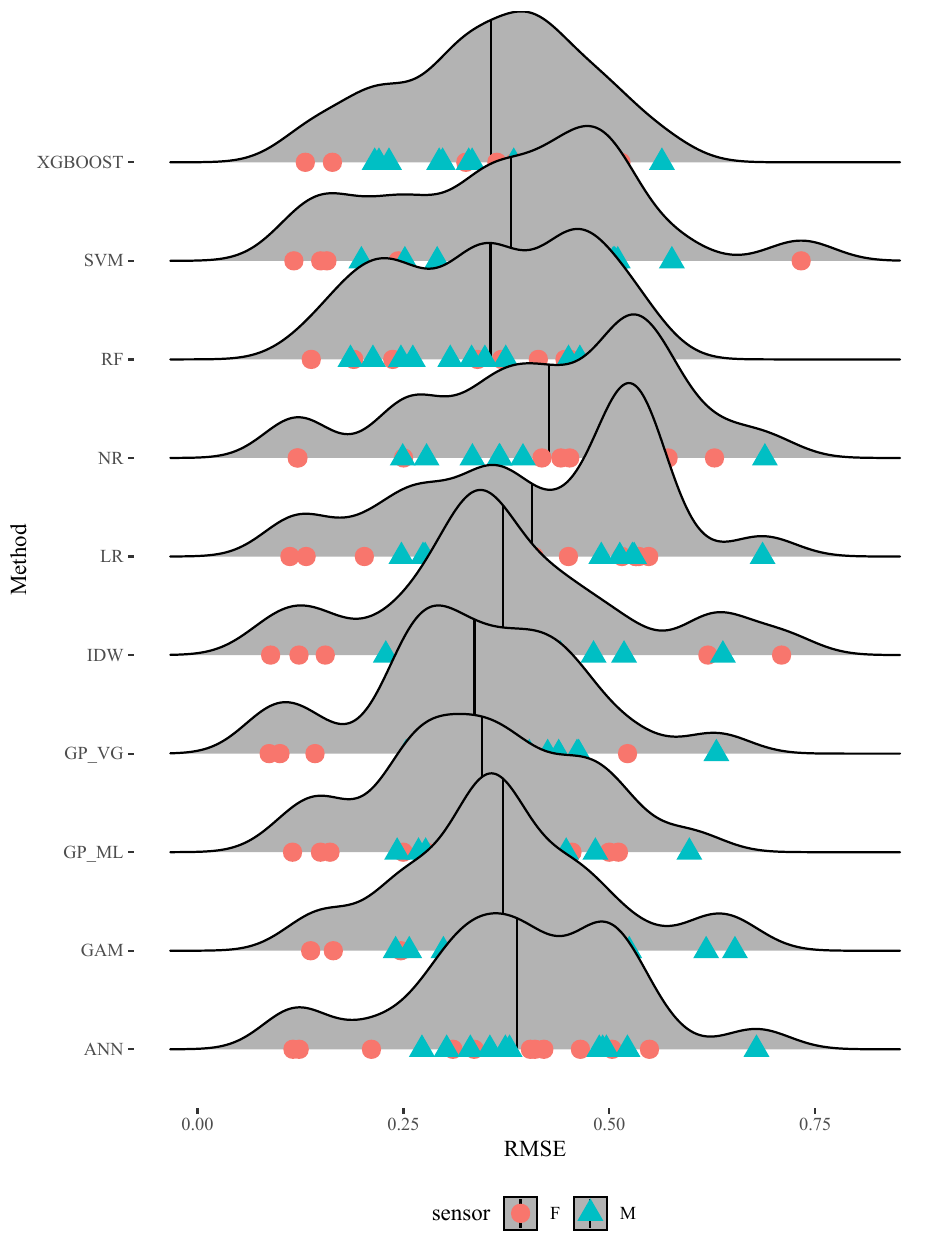}
\caption{RMSE VC}
\label{VC_lcs}
\end{figure}

Starting by looking at the RMSE of the 10 models in the simplest scenario, temporal interpolation in figure \ref{VC_lcs}, we notice that all the models present similar results with an average RMSE around 0.37 log(µg)/m3. 
With a minimum of 0.087 log(µg)/m3 obtained by the Gaussian process model estimated by space-time variograms. The maximum is found in the SVM model with 0.733 log(µg)/m3.
We also observe the similarity of the two RMSE distributions for the Gaussian process estimated by maximum likelihood and least-squares variograms.
Similarly for the RMSE distribution of the simple linear model and network regression, given the similarity of the two models, we note a better score for multiple linear regression.
Due to the proximity of some low-cost fixed sensors, the prediction for these sensors becomes easy for all studied models.


\begin{figure}[h]
\includegraphics[width=\columnwidth]{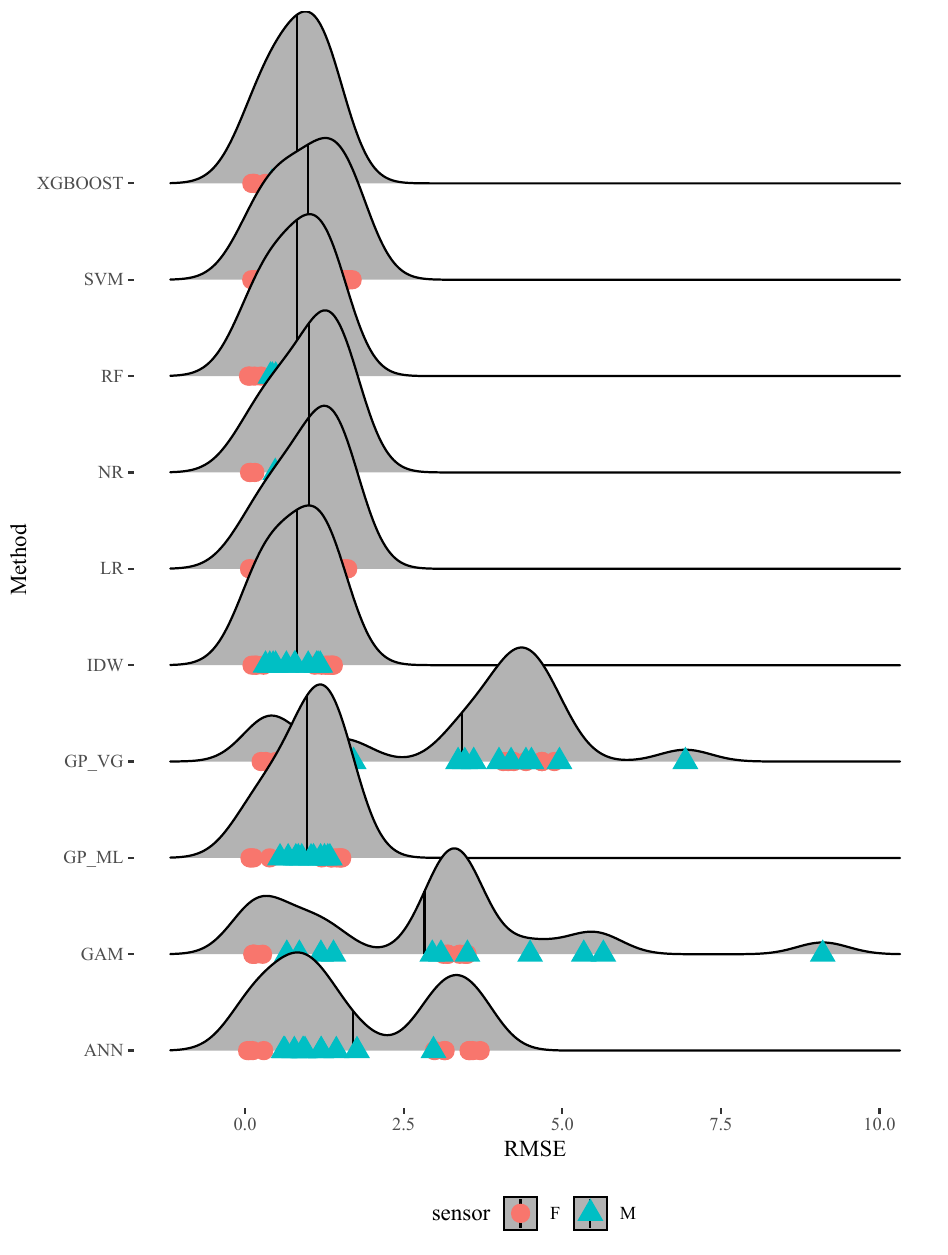}
\caption{Correlation on LCS (prevision scenario)} 
\label{prev_lcs}
\end{figure}

Figure \ref{prev_lcs} shows that the error is larger when trying to predict in a day when we have no observation, more precisely, the average RMSE of the 10 models is 1.44 log(µg)/m3.
The minimum of 0.87 log(µg)/m3 is obtained by the XGBOOST algorithms. The maximum is found in the Geostatistical Model with a spatio-temporal variograms model with 3.42 log(µg)/m3.

Unlike the previous scenario, we obtain a more accurate prediction: in the forecast scenario, the most important are the covariates, since the observations are temporally far, knowing that the mobile sensors have more varying covariates, we can explain more of the variance on the mobile sensors.

While LR and GPM maintain the same error distribution, a big difference appears between the two ways of estimating the parameters of the Gaussian process. The maximum likelihood method gives better results for two reasons; the first is that the covariate parameters are better estimated (taking into account the dependence of the observations in the case of the maximum likelihood method), the second reason is related to the way of estimating the covariate parameters in the variogram model. We set a higher weight at the origin of the variogram, and so we predict better for close observations.
We also notice that the highly complex models (ANN, GAM) are not efficient in the case of unsampled days forecasts.

\subsubsection{Validation on fixed monitoring stations}

\begin{figure}[H]
\includegraphics[width=\columnwidth]{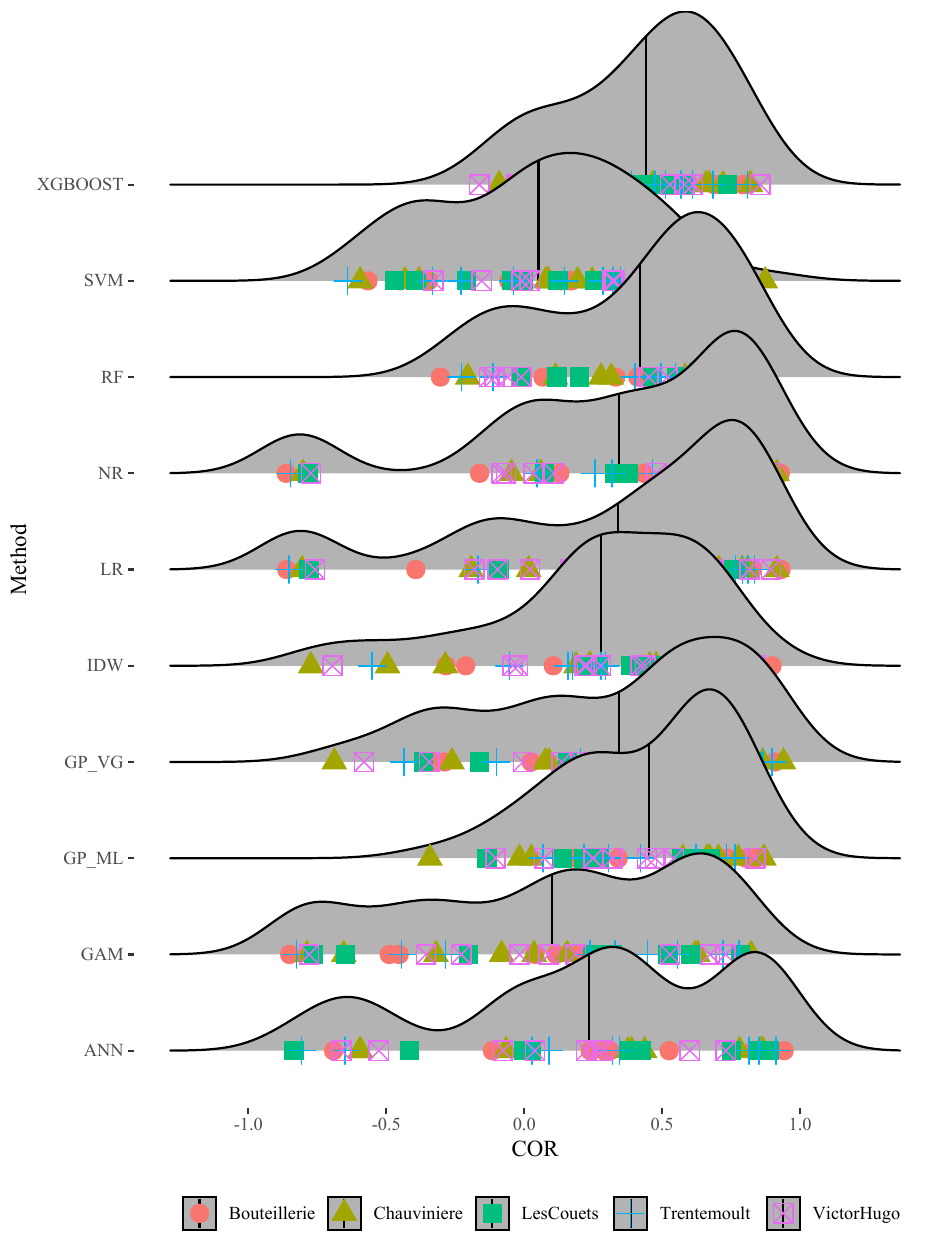}
\caption{Correlation on fixed monitoring stations in the first scenario} 
\label{VC_fixe}
\end{figure}
Focusing on the correlations of the predictions on the fixed air quality stations, we notice that in the temporal interpolation scenario represented in Figure \ref{VC_fixe}, there is a negative correlation on some stations for all models.
On average for all models, we obtain a correlation score of 0.3.
The maximum likelihood Gaussian process model has the best score in terms of average correlation on all stations with 0.45, against 0.05 for SVM which records the lowest score.

\begin{figure}[H]
\includegraphics[width=\columnwidth]{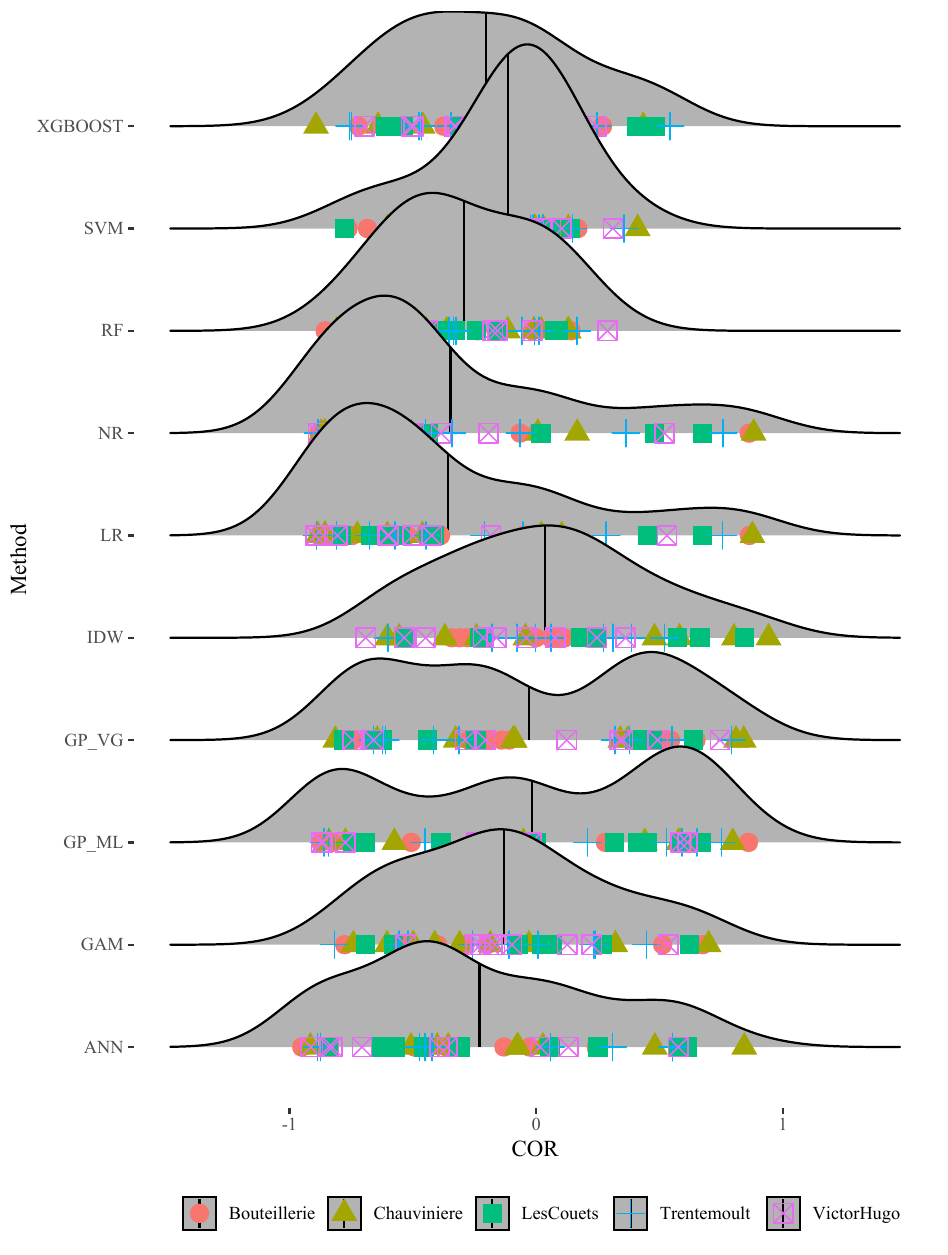}
\caption{Correlation on fixed monitoring stations in the second scenario} 
\label{prev_fixe}
\end{figure}

The same observation as with the cross-validation on low-cost sensors is that as soon as we are interested in a prediction, most of the models fail to give better results. This is even worse in the case of a validation on fixed air quality stations, where we can obtain a negative mean correlation for all models except IDW.
The average correlation for all models is -0.16.
We deduce from Figure \ref{prev_fixe} the inability of the models to predict the measurements collected by the reference stations using earlier data.

\subsubsection{Summary table }
The Table \ref{Summary table} shows all the performance indicators results for all models in the two scenarios and the two validation methods.
In the case of a cross-validation on low-cost sensors, all models are roughly equivalent and offer excellent results, with equivalent RMSE as seen in Figure \ref{VC_lcs} and equivalent correlations ranging from 0.75 to 0.86. Moreover, in this case, neither the bias nor the MAE points to a faulty model.

In the case of low-cost sensor forecasting, the RMSE scores drop for all models as seen in Figure \ref{prev_lcs}, as well as for the correlations that now range from 0.15 to 0.43.
Examination of the bias as well as MAE concludes to the inadaptation of the models ANN, GAM, and GPVG in the case of a forecast.

Concerning the validation on fixed air quality stations, the RF, XGBOOST, and ML\textunderscore GP models offer the best results on all performance indicators, with correlations greater than 0.4 for these three models.
We also note that the GAM and ANN models perform significantly worse.

Lastly, when looking at air quality station forecasts, the study of performance indicators indicates that all models are inadequate.

Overall, machine learning models except for ANN (XGBOOST RF and SVM) seem more adapted to the forecast scenario.
As expected, we make more errors in the prevision scenario, whether validated on LCS or fixed stations.
We make many more errors than we validate on the fixed stations, which can indicate a difference in the nature of the data between that collected by the LCS and that of the fixed stations.

\begin{table*}[h!]
\begin{center}

\begin{tabular}{|c|c|c|c|c|c|c|}
\hline
Model & Scenario & Validation & RMSE  & COR & BIAS &  MAE              \\ \hline \hline
\multirow{4} {*} {IDW}   & \multicolumn{1}{c|}{\multirow{2}{*}{Interpolation}} &Low cost sensors & 0.371 & 0.826 &-0.005&2.29\\ \cline{3-7} 
                             &                                               &Fixed monitoring station & 0.760 &0.277  &  0.0403 &1.56  \\ \cline{2-7} 
                             & \multirow{2}{*}{Prevision}                       &Low cost sensors & 0.822 &0.353&-0.454&3.37 \\\cline{3-7} 
                             &                                               & Fixed monitoring station  & 0.892  & 0.0366& -0.434&1.53  \\ \cline{1-7}
\multirow{4} {*} {LR}   & \multicolumn{1}{c|}{\multirow{2}{*}{Interpolation}} & Low cost sensors & 0.406 & 0.766& 0.018&2.25\\ \cline{3-7} 
                             &                                               & Fixed monitoring station  & 0.915 &0.340&-0.204 & 1.71  \\ \cline{2-7} 
                             & \multirow{2}{*}{Prevision}                       &Low cost sensors  &  1.00   &0.409&-0.219& 3.78 \\\cline{3-7} 
                             &                                               & Fixed monitoring station  & 1.67  &-0.356&-0.513&2.70  \\ \cline{1-7}
\multirow{4} {*} {NR}   & \multicolumn{1}{c|}{\multirow{2}{*}{Interpolation}} &Low cost sensors  & 0.427& 0.759 &0.023&2.32\\ \cline{3-7} 
                             &                                               & Fixed monitoring station  &  0.990 & 0.343 &-0.239& 1.73  \\ \cline{2-7} 
                             & \multirow{2}{*}{Prevision}                       & Low cost sensors & 1.01  &0.411& -0.229&3.77 \\\cline{3-7} 
                             &                                               & Fixed monitoring station  &  1.69 &-0.347&-0.502& 2.75  \\ \cline{1-7}
\multirow{4} {*} {GAM}   & \multicolumn{1}{c|}{\multirow{2}{*}{Interpolation}} & Low cost sensors  & 0.371& 0.840&-0.066&2.22\\ \cline{3-7} 
                             &                                               &Fixed monitoring station  & 4.58 & 0.0996& -2.72 &8.74 \\ \cline{2-7} 
                             & \multirow{2}{*}{Prevision}                       & Low cost sensors  & 2.83  &0.243&0.886 &11.9 \\\cline{3-7} 
                             &                                               & Fixed monitoring station  &  3.43  &-0.130& -1.126& 6.39  \\ \cline{1-7}
\multirow{4} {*} {GSTAT}   & \multicolumn{1}{c|}{\multirow{2}{*}{Interpolation}} & Low cost sensors &  0.336&0.840&0.007&2.18\\ \cline{3-7} 
                             &                                               & Fixed monitoring station  &  1.18  & 0.343 &-0.456 &2.58  \\ \cline{2-7} 
                             & \multirow{2}{*}{Prevision}                       & Low cost sensors  & 3.42  &0.201&0.0242&11.8 \\\cline{3-7} 
                             &                                               & Fixed monitoring station &3.62  &-0.0293& -0.518& 6.71  \\ \cline{1-7}
\multirow{4} {*} {GPGP}   & \multicolumn{1}{c|}{\multirow{2}{*}{Interpolation}} & Low cost sensors  &0.345& 0.810&0.013&2.13\\ \cline{3-7} 
                             &                                               & Fixed monitoring station  & 0.863 &0.452& -0.193 &1.60  \\ \cline{2-7} 
                             & \multirow{2}{*}{Prevision}                       & Low cost sensors  & 0.980  &0.368& -0.356 &3.08 \\\cline{3-7} 
                             &                                               &Fixed monitoring station  &    1.17   & -0.015&-0.642&1.95  \\ \cline{1-7}
\multirow{4} {*} {RF}   & \multicolumn{1}{c|}{\multirow{2}{*}{Interpolation}} & Low cost sensors  & 0.356& 0.852&-0.097& 2.22\\ \cline{3-7} 
                             &                                               &Fixed monitoring station  & 0.677 &0.419&-0.103 & 1.31  \\ \cline{2-7} 
                             & \multirow{2}{*}{Prevision}                       & Low cost sensors &  0.824 &0.409&-0.362 &3.30 \\\cline{3-7} 
                             &                                               & Fixed monitoring station  &0.873 &-0.291& -0.299& 1.55 \\ \cline{1-7}
\multirow{4} {*} {ANN}   & \multicolumn{1}{c|}{\multirow{2}{*}{Interpolation}} &Low cost sensors  &0.388& 0.756&-0.0574&2.32\\ \cline{3-7} 
                             &                                               & Fixed monitoring station  &1.65 & 0.234&-0.634&2.84  \\ \cline{2-7} 
                             & \multirow{2}{*}{Prevision}                       & Low cost sensors  &1.71   &0.152&-0.626&9.11 \\\cline{3-7} 
                             &                                               & Fixed monitoring station  & 2.01  & -0.230& -1.33&4.06  \\ \cline{1-7}
\multirow{4} {*} {SVM}   & \multicolumn{1}{c|}{\multirow{2}{*}{Interpolation}} & Low cost sensors  &0.381& 0.802&-0.058&2.26\\ \cline{3-7} 
                             &                                               & Fixed monitoring station  &  0.750 &0.0518&-0.125&1.38  \\ \cline{2-7} 
                             & \multirow{2}{*}{Prevision}                       & Low cost sensors  & 0.993  &0.351&-0.555 &3.65 \\\cline{3-7} 
                             &                                               & Fixed monitoring station  &0.817  &-0.113& -0.339&1.53  \\ \cline{1-7}
\multirow{4} {*} {XGBOOST}   & \multicolumn{1}{c|}{\multirow{2}{*}{Interpolation}} & Low cost sensors  & 0.356& 0.861&-0.0894& 2.27\\ \cline{3-7} 
                             &                                               & Fixed monitoring station  & 0.709 &0.443&-0.113& 1.47  \\ \cline{2-7} 
                             & \multirow{2}{*}{Prevision}                       & Low cost sensors  &  0.814  &0.429&-0.357 &3.34\\\cline{3-7} 
                             &                                               & Fixed monitoring station  & 0.876  &-0.202&-0.290&1.60  \\ \cline{1-7}
                             
\end{tabular}
\end{center}
\caption{Summary table} 
\label{Summary table}
\end{table*}

\subsection{Map}

\begin{figure*}[h!]
\includegraphics[height=7cm,width=\textwidth]{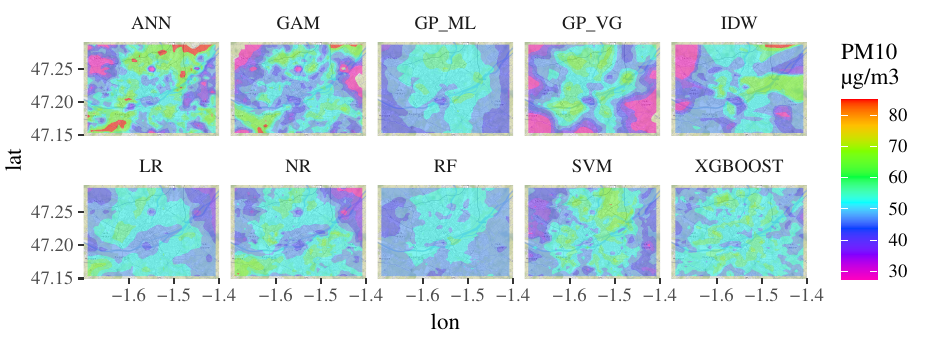}
\caption{Maps produced by the different models} 
\label{MAP}
\end{figure*}

Figure \ref{MAP} displays the maps generated by 10 different models at 12:00 on 26/11/2018, utilizing all the data for that day. We selected this day because it had the highest amount of data, exceeding 19500 observations.

As anticipated, the maps produced by the linear and network regression models are almost indistinguishable. We also notice similarities between the two geostatistics models, but the differences become more apparent towards the map's borders as we move away from the observations.

For all models except IDW, we observe a tendency of areas of pollutant concentration around the center and periphery of the road in Nantes city. Meanwhile, lower concentration values are present on Nantes Island, where most of the fixed LCS sensors are located.

It is important to note that these maps only illustrate the spatial dispersion of concentrations at a specific time of the day and not their temporal evolution. It is also known that data have a greater variance over time than in space.

\section{Discussion}

Simple versions of the models have been used, and one can increase the complexity of each of the models presented here. GAM by making relations, geostatistics by modeling the anisotropy, or neural networks by increasing the number of layers used. However, we have chosen to do this to give a baseline of the results for the models.

The mobile LCS only measured one kind of urban pollution because it did not cross all types of streets. More precisely, as it is an opportunistic sampling, it is more likely to sample the main traffic arteries and measure the pollution due to traffic and not the background pollution captured by some fixed sensors, LCS or reference stations.

In addition, the data from the mobile sensors were considered independent of the concentration of PM, which means that this paper did not take into account the autocorrelation of the data from the same sensor.

The study of the relationship between the ratio of fixed and mobile LCSs to prediction accuracy is complex. The data from the mobile sensor are of lower quality but greater quantity. Moreover, the probability that a sensor fails during a mobile run is greater than for the fixed ones. These factors are difficult to take into account for an objective numerical comparison.

The reduction of standard errors in a Gaussian process, which does not even require data after the parameter estimation step, can be used to study the theoretical impact of the fixed/mobile ratio on prediction accuracy.
It was the procedure followed by \cite{guan2020fine}, and this approach has several limitations. First, the model is an approximation of the real dispersions of the physical process. Concluding on the optimal configuration of the sensors from the characteristics of a reduced model might be misleading.
Secondly, it does not take into account the contribution of the sensors to the estimation of the parameter step, specifically, the mean parameters.

If we do not use any mobile sensor, we can even fall into the second case of the problem of identifiability of the parameters for spatial covariates \cite{guillaume2019introductory}.

Two aspects of the issue must be distinguished: for parameter estimations, more mobile sensors will be preferred. For prediction accuracy, i.e. minimize the errors of prediction, we are not able to answer this question although the fixed sensors will have an advantage because they require less effort for it to work continuously.

Overall, choosing between a mobile or fixed sensor is challenging and depends enormously on the sensor and the model, it is also related to their use.

The quality of the data is more important than the statistical method. It is not recommended to use LCS alone; instead, they should always be systematically coupled with more reliable data.




The inclusion of CHIMERE outputs in the comparison could have been considered; however, due to disparities in the input data, such a comparison was inappropriate. It is important to note that the second scenario incorporates the forecast as a covariate, whereas the first scenario employs the analysis as a covariate.

The way of estimating the parameters of the same model has a real impact on the obtained results, the GP\textunderscore VG model focuses on the estimation of the parameters at the beginning of the variogram, putting the emphasis on "close" predictions. 


\section{Conclusion}

In conclusion, this study demonstrates the feasibility and potential of using mobile sensors to map air quality in urban areas. 
This article reviews ten models from the literature using a low-cost fixed and mobile sensor data set together with external variables in the city of Nantes. We compare them on the basis of predefined performance indicators estimated from an experimental dataset. We show the existence of a significant bias between the output of the models using low-cost sensor data and the validation data of the fixed air quality stations considered as reference, in a time interpolation and time prediction scenario. We also show that machine learning models are more suitable in the forecasting scenario. The results show that the RF, XGBOOST and ML\textunderscore GP models offer the best results on all performance indicators, with correlations greater than 0.4 for these three models.
The quality of the data is more important than the statistical method. It is not recommended to use LCS alone; instead, they should always be systematically coupled with more reliable data.
\section{Aknowledgements}
The authors thank the Atmotrack company (Nantes, France) for providing the full data set that was used in this article.

\bibliographystyle{apalike}
\bibliography{bibfile}

\end{document}